\documentclass[aps,epsfig,letter,nofootinbib,superscriptaddress,twocolumn,floatfix]{revtex4} 
\usepackage[dvips]{graphicx} 
\usepackage{amssymb}
\usepackage[usenames]{color}
\newcommand{\sfig}[2]{
\centerline{ \includegraphics[width=#2]{#1} }
		}
\newcommand{\Sfig}[2]{
	\begin{figure}[thp]
	\sfig{#1.eps}{0.9\columnwidth}
	\caption{{\small #2}}
	\label{fig:#1}
	\end{figure}
}

\newcommand{\rf}[1]{\ref{fig:#1}}

\newcommand{\be}{\begin{equation}}
\newcommand{\ee}{\end{equation}}
\newcommand{\bea}{\begin{eqnarray}}
\newcommand{\eea}{\end{eqnarray}}
\newcommand{\eql}[1]{\label{eq:#1}}       
\newcommand{\refeq}[1]{Eq.~(\ref{eq:#1})}

\newcommand{\reffig}[1]{Fig.~\ref{fig:#1}}

\newcommand{\secspace}{-0.5cm}
\newcommand{\sectspace}{-0.3cm}

\renewcommand{\v}[1]{\vec{#1}}

\renewcommand{\k}{\kappa}

\renewcommand{\l}{\ell}
\newcommand{\vth}{\v{\theta}}
\newcommand{\eps}{\varepsilon}

\newcommand{\iab}{i_{AB}}
\newcommand{\ri}{r}

\newcommand{\rcut}{r_{\rm cut}}

\begin{document}

\title{Size Bias in Galaxy Surveys}

\author{Fabian Schmidt}
\affiliation{Department of Astronomy \& Astrophysics, The University of
Chicago, Chicago, IL 60637-1433}
\affiliation{Kavli Institute for Cosmological Physics, Chicago, IL 
60637-1433}

\author{Eduardo Rozo}
\affiliation{CCAPP, Ohio State Universtiy, Columbus, OH 43210}

\author{Scott Dodelson}
\affiliation{Center for Particle Astrophysics, Fermi National Accelerator 
Laboratory, Batavia, IL 60510-0500}
\affiliation{Department of Astronomy \& Astrophysics, The University of
Chicago, Chicago, IL 60637-1433}
\affiliation{Kavli Institute for Cosmological Physics, Chicago, IL 
60637-1433}

\author{Lam Hui}
\affiliation{Department of Physics, ISCAP, Columbia University, New York, NY 10027}

\author{Erin Sheldon}
\affiliation{Brookhaven National Laboratory, Upton, NY 11973}

\begin{abstract}
Only certain galaxies are included in surveys: those bright and large enough to be detectable as extended sources. 
Because gravitational lensing can make 
galaxies appear both brighter and larger, the presence of foreground inhomogeneities can scatter 
galaxies across not only magnitude cuts but also size cuts,  changing the statistical properties of the resulting catalog. Here 
we explore this size bias, and how it combines with magnification bias to affect galaxy statistics. 
We demonstrate that photometric galaxy samples from current and upcoming surveys can be even more 
affected by size bias than by magnification bias. 
\end{abstract}

\date{\today}

\maketitle

\section{Introduction}
\vspace*{\sectspace}

Any survey using either counts of galaxies or shapes of galaxies must make choices: which galaxies are in and which are out. For example, a magnitude limited survey contains all galaxies brighter than a fixed flux threshold. Because gravitational lensing can make galaxies appear brighter, a background galaxy whose line of sight passes through an overdense region might be catapulted into the sample when it otherwise would not have been included in the catalog.   As was pointed out many years ago~\cite{tog,whhw,fug88,nar89,sch89,Matsubara2000}, this leads to non-zero correlations between background and foreground objects, an effect which goes by the monicker of {\it magnification bias} \cite{VallinottoEtal2007,HuiGaztanagaLoVerde2007,SchmidtEtal}.

A less studied phenomenon is that the increase in {\it size} of a background galaxy by a foreground matter overdensity \cite{jain02}
will have a similar effect if the catalog under consideration includes a size cut.  We refer to this additional source of bias as {\it size bias}, and to the net effect of gravitational lensing
on the galaxy sample as {\it lensing bias}.

The purpose of this paper is to introduce size bias as an additional 
effect that needs to be considered in studies of galaxy statistics.  
Any study which correlates galaxies in a flux- and size-limited sample 
with some other observable is
potentially sensitive to size as well as magnification bias.
In addition, size and magnification bias affect {\it shear} observables as well.
This effect has so far not been studied in detail and is fleshed out in 
a companion paper \cite{paperII}.


\Sfig{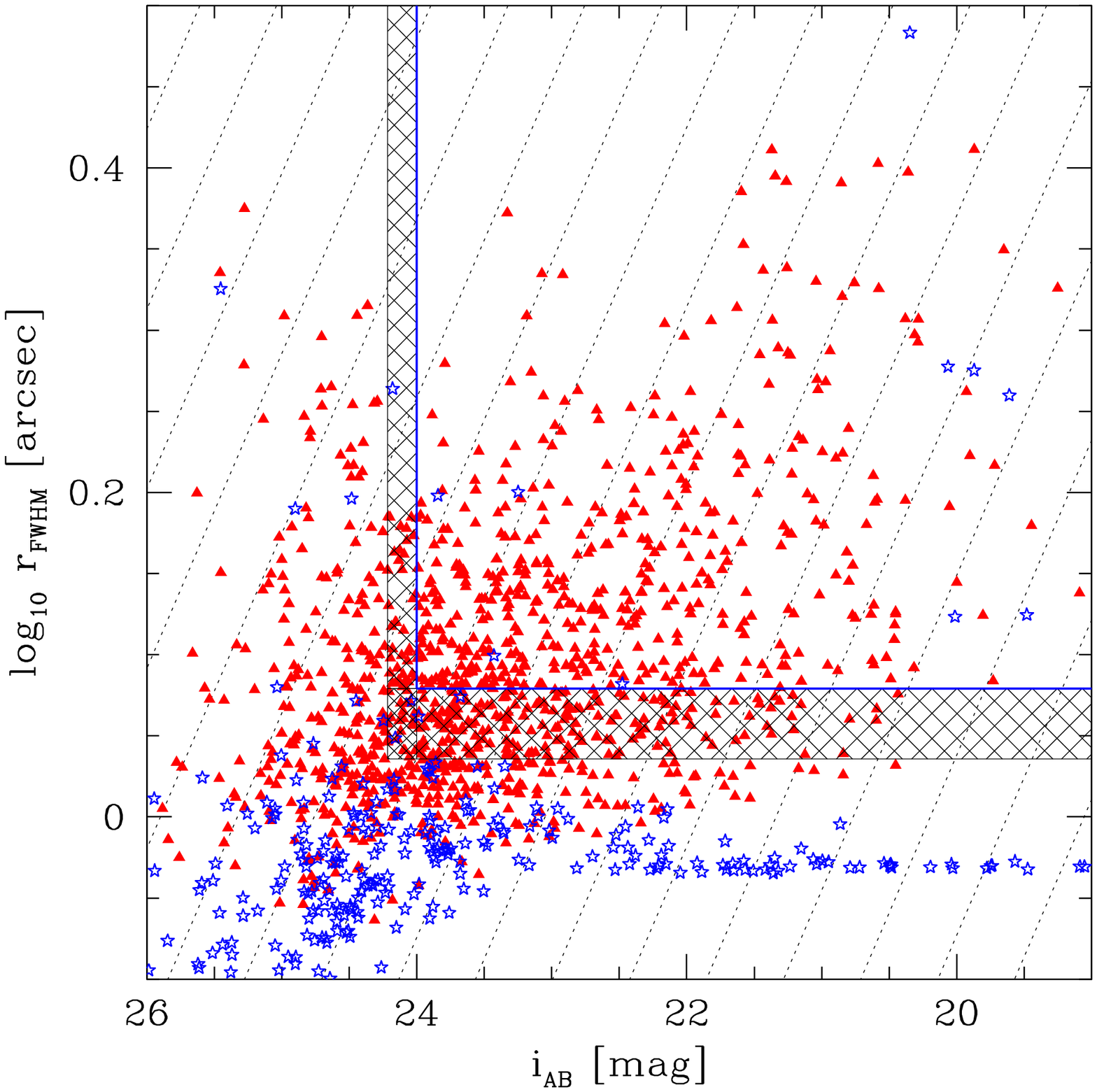}{The distribution of galaxies (red triangles) and stars 
(blue) in the GOODS field as a function of size and magnitude, after downgrading images
to an assumed $0.9"$ seeing, typical of near-future large imaging surveys such as DES and Pan-STARRS. The thick vertical and horizontal lines denote our fiducial magnitude and size cuts
($i_{AB}<24$ and $r > 1.2$~arcsec, respectively). The hatched regions 
contain galaxies included in the survey due to lensing by a foreground mass:
the vertical band corresponds to the traditional magnification bias effect, 
while the horizontal region contains galaxies affected by the size bias 
introduced in this work. ``Stars'' at lower left with abnormally small sizes are shot noise.}


\vspace*{\secspace}
\section{Lensing Bias}
\vspace*{\sectspace}

\textit{Qualitative discussion.} Figure
\rf{flux-size} illustrates the basic physics behind lensing bias.  The figure
shows size vs.  $i$ band apparent magnitude for a small random subsample of
stars and galaxies in the GOODS field~\cite{:2003ig}.  
Our size variable is $\ri$, the
Full Width at Half Maximum (FWHM) of the galaxies in the $i$ 
band.  The FWHM
was measured using adaptive moments \citep{BJ02,HirataSeljak03}. All images
have been degraded by adding noise and $0.9"$ seeing typical of the upcoming
Dark Energy Survey \citep[DES,][]{DES05}.
Also shown in the
figure are horizontal and vertical blue lines corresponding to a size and
magnitude cut of $\ri> 1.2"$ and $\iab<24$, respectively.  The magnitude cut is
necessary to obtain high S/N images in surveys with DES-like depth, while the
size cut is meant to be comfortably larger than the $0.9"$ seeing of the
experiment, which is necessary in order to be able to robustly separate
galaxies from stars, and to obtain good shape measurements of the 
galaxies\footnote{For faint objects in the 
downgraded data, photon noise leads to 
large scatter in the estimated $r_{\rm FWHM}$. Note that the star-galaxy 
separation is technically done based on the resolution parameter $R$ 
\cite{BJ02}, not $r_{\rm FWHM}$, and the relation between the two has some 
scatter.}.

The diagonal dotted lines in Figure \rf{flux-size} are contours of constant surface brightness.  Since lensing preserves surface brightness, a foreground matter overdensity will move all sources along these diagonal lines upwards and to the right, increasing the number of objects in the galaxy sample. 
Conversely, an underdense region along the line of sight can act in the
opposite way.
In the weak lensing limit, the flux magnification factor $A$ is given by
$A= 1+2\kappa$, where $\kappa$ is the convergence (see below). 
Then, the magnitude and 
size of a source will be (de-)magnified according to:
%
\begin{equation}
\iab \rightarrow\iab - \frac{5\kappa}{\ln 10};\ \ 
\lg \ri \rightarrow \lg \ri + \frac{\kappa}{\ln 10}. \label{eq:scaling}
\end{equation} 
The narrow horizontal and vertical hatched bands shown in Figure~\rf{flux-size} contain galaxies which would pass our galaxy selection cuts if they were to be lensed by matter along the line of sight. For illustration purposes, we chose a
large value of the lensing convergence, $\kappa=0.1$.  The well known 
magnification bias effect corresponds to the vertical hatched region.  
In addition, however, we see that there is a comparable population of galaxies 
in the horizontal hatched region which is brought into our galaxy sample not 
because the galaxies are made brighter, but because the galaxies become more 
extended.  

\textit{Quantitative treatment.} We wish to determine how the galaxy density field $n_{\rm obs}(\vth)$ observed in an experiment relates to the intrinsic (i.e. un-lensed)
galaxy density field $n_0(\vth)$.   Our discussion follows closely
the presentation found in the appendix of \citet{Huietal07}.
Let $f$, $r$, and $\vth$ denote the observed flux, apparent size, and position
of a galaxy, respectively, and $f_g$, $r_g$, and $\vth_g$ denote the corresponding intrinsic
quantities. Further, let $\Phi(f,r,\vth)$ and $\Phi_g(f_g,r_g,\vth_g)$ denote the 
observed and intrinsic distribution of galaxies in flux, size, and position. 
Because the total number of galaxies is preserved, one has that
\be
\Phi(f,r,\vth)\:df\: dr\: d^2\vth = \Phi_g(f_g,r_g,\vth_g)\: df_g\: dr_g\: d^2\vth_g.
\label{eq:galconserve}
\ee

In the linear regime, the intrinsic and observed galaxy properties are related via\footnote{Linear distortions of galaxies adequately describe gravitational 
lensing everywhere except where galaxies fall near the caustic of a foreground lens.  Also, while the impact of gravitational lensing on the source radius
$r$ may depend on the precise definition of $r$, we expect the relation $r=A^{1/2}r_g$ to hold in the vast majority of cases to a good approximation.  Other scalings
can be trivially included following the discussion presented in this work.  } 
\be
\vth=\vth_g+\delta\vth,\ \  f = A\:f_g,	\ \   r=\sqrt{A}\:r_g,\ \  d^2\vth = A\, d^2\vth_g
\label{eq:lensing}
\ee
where $\delta\vth$ is the gravitational lensing deflection angle, and $A=|\partial \vth/\partial \vth_g|$ is the corresponding amplification.  
We work in the weak lensing regime throughout, so that
$A=1+2\kappa$. Note that the surface brightness of a source
$S \propto f/r^2$ is unaffected by $A$.

Let now $\eps(f,r)$ be the galaxy selection function of a survey, which we assume depends only on a source's magnitude and size.
Given this selection function, the intrinsic and observed galaxy number
densities are:
\bea
n_{\rm obs}(\vth) &=& \int df \int dr\: \eps(f,r) \Phi(f,r,\vth),\\
n_0(\vth_g) &=& \int df_g \int dr_g\: \eps(f_g,r_g) \Phi_g(f_g,r_g,\vth_g).
\eea
Using equation (\ref{eq:lensing}), we can write $f,r,$ and $\vth$ in the above expression in terms of the corresponding intrinsic quantities.
Linearizing, and using equation (\ref{eq:galconserve}) to replace
$\Phi$ by $\Phi_g$, we find that the relation between the observed and intrinsic galaxy densities is given by 
\be
n_{\rm obs}(\vth) = n_0(\vth) [ 1 + (2\beta_f+\beta_r - 2)\k ],\label{eq:nobs}
\ee
where the parameters $\beta_f$ and $\beta_r$ are defined via: 
\bea 
\beta_f &\equiv& \frac{1}{n_{\rm obs}}\int dr\int df \frac{\partial\eps(f,r)}{\partial \ln(f)} \Phi(f,r) \label{eq:betaf}\\
\beta_r &\equiv& \frac{1}{n_{\rm obs}}\int dr\int df \frac{\partial\eps(f,r)}{\partial \ln(r)} \Phi(f,r).\label{eq:betas}
\eea
While strictly speaking $\beta_f$, $\beta_r$ are defined in terms of the
{\it intrinsic} galaxy distribution $\Phi_g$, we have replaced this with 
the {\it observed} galaxy distribution $\Phi$, which is the only one accessible
to observations. Any differences between the two are of order
$\kappa$ averaged over the survey area, and hence correspond to a higher order
correction in our perturbative approach.

If we further assume that the selection function is given by a simple magnitude and size cut, the function $\eps(f,r)$ takes the form 
$\eps(f,r) = \Theta(f-f_{\rm min})\Theta(r-r_{\rm min})$ where $\Theta(x)$ is a step function.  In this case, the expressions for
$\beta_f$ and $\beta_r$ simplify to 
%
\be
\beta_f = -\frac{\partial \ln n_{\rm obs}}{\partial \ln f}\Bigg\vert_{\stackrel{f=f_{\rm min}}{r=r_{\rm min}}}; \ \  \beta_r = -\frac{\partial \ln n_{\rm obs}}{\partial \ln r}\Bigg\vert_{\stackrel{f=f_{\rm min}}{r=r_{\rm min}}}.
\label{eq:finalb}
\ee
Note that in this case, one has $\beta_f = (5/2) \partial \lg n / \partial m \equiv 5s/2$.  
If $\beta_r=0$, one obtains $n_{\rm obs} = n_0 [1+(5s-2)\kappa]$, which
is the standard expression for magnification bias. For $\beta_r\neq 0$, size 
bias acts in a way that is analogous to magnification bias, except that it is 
sensitive to the slope of the galaxy size distribution rather than that of
the magnitude distribution.
The parameter which captures the full extent of lensing bias therefore is 
\be
q \equiv 2\beta_f+\beta_r-2.
\ee
To first order in $\kappa$, lensing bias affects the observed number density via
\be n_{\rm obs}=n_0(1+q\kappa)\eql{q}.
\label{eq:nobss}
\ee  
The next section presents estimates of $q$ for upcoming galaxy surveys. 


\begin{figure}[t!]
\centering
\includegraphics[width=0.48\textwidth]{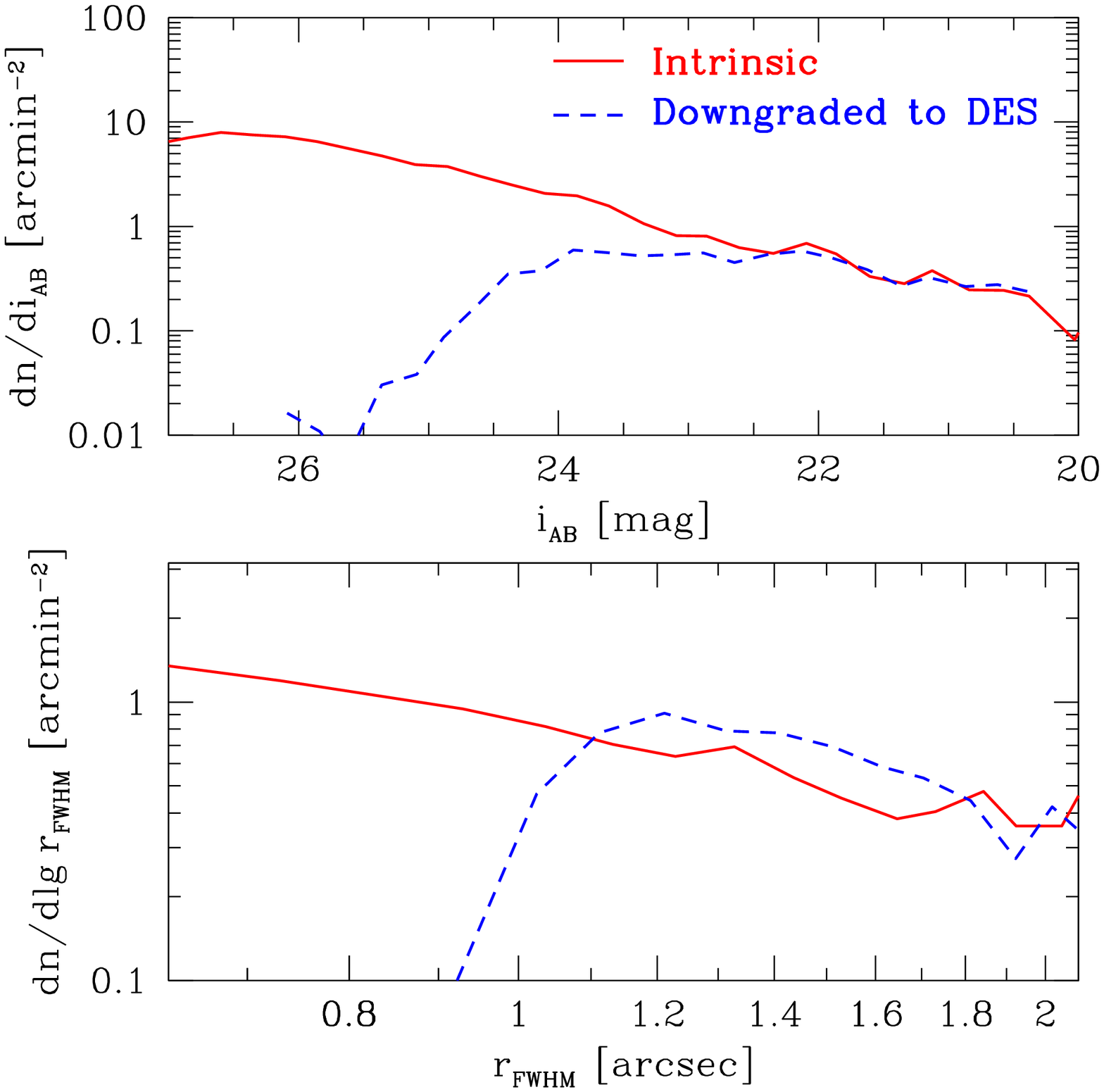}
\caption{{\small Magnitude ($\iab$, \textit{top panel}) and apparent size 
($\lg r$, \textit{bottom panel}) distribution  of galaxies in the 
GOODS field. 
In each case, the solid red line shows the full-resolution GOODS data,
while the blue dashed line shows the distribution for data degraded to the expected
DES instrument, typical noise, and $0.9"$ seeing.
}
\label{fig:cuts}}
\end{figure}


\begin{figure}[t!]
\centering
\includegraphics[width=0.48\textwidth]{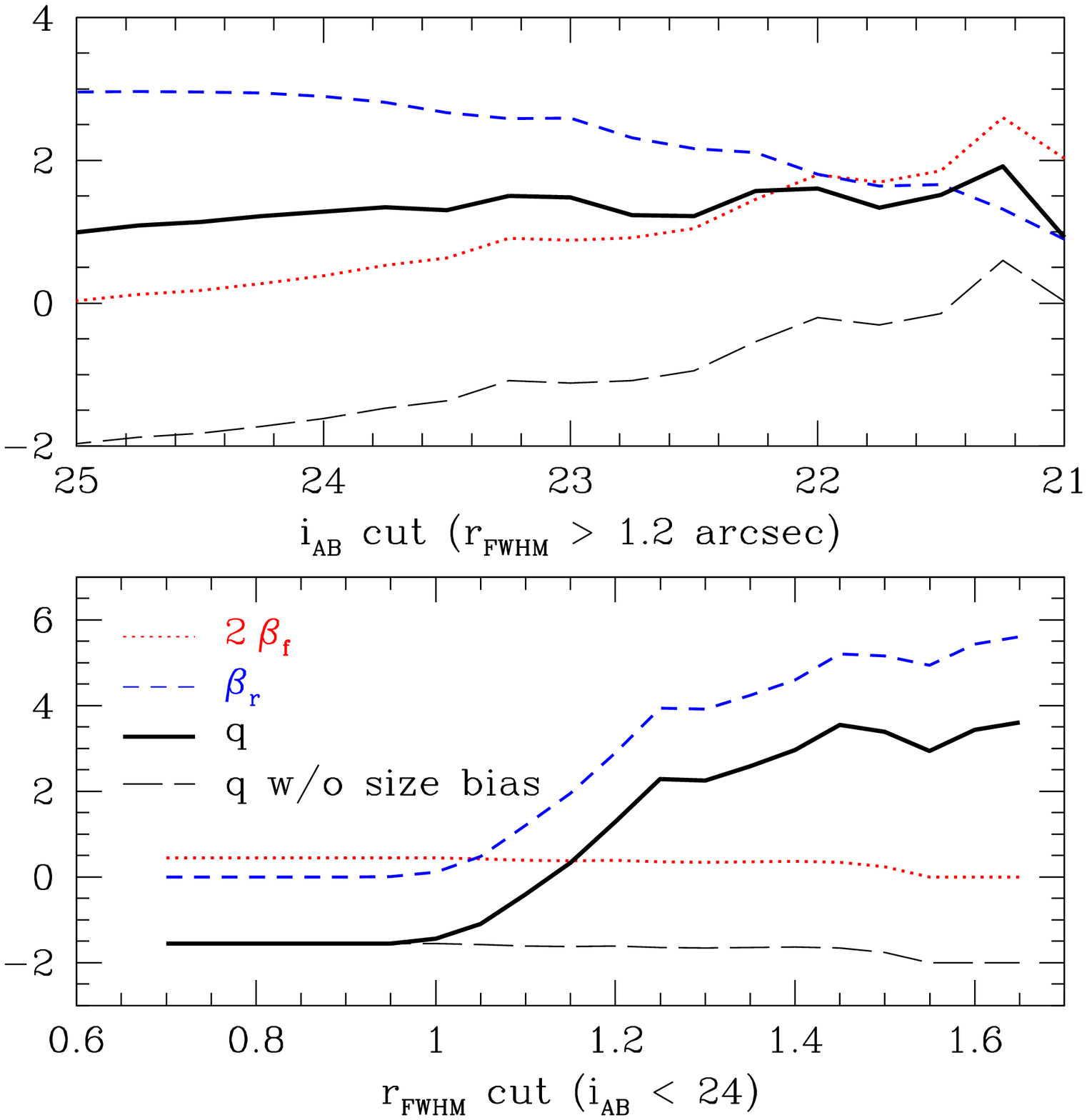}
\caption{{\small Flux- and size count slopes,
$2\beta_f$ and $\beta_r$, measured from GOODS data degraded to DES
seeing and noise, as a function of the magnitude cut for a fixed size cut
($r > 1.2$~arcsec; \textit{top panel}), and as a function
of the size cut for a fixed magnitude cut ($i_{AB} < 24$~mag; 
\textit{bottom panel}). Also shown is the total $q$ from \refeq{nobss} 
(solid line), and the $q$ obtained when neglecting size bias (long-dashed line).}
\label{fig:beta}}
\end{figure}


\vspace*{\secspace}
\section{Forecast for upcoming surveys}
\label{sec:forecast}
\vspace*{\sectspace}

Figure \ref{fig:cuts} shows the galaxy density of the
GOODS field as a function of both magnitude and size.  The GOODS data is
degraded to $0.9"$ seeing and typical noise, and the galaxies are error
weighted using adaptive moments \citep{BJ02,HirataSeljak03}.  In order for
galaxy ellipticities to be reliably measured, one requires both $\iab \lesssim
24$ and $r \gtrsim 1.2"$, which we adopt as our fiducial magnitude and size
cuts.  

In order to measure the $q$ parameter for this data set, we
determine $\beta_f$, $\beta_r$ from the downgraded GOODS galaxy sample
according to \refeq{finalb}, for a range of
cut values.
Note that,
consistent with our perturbative approach, we have neglected the intersection 
between the two hatched regions in \reffig{flux-size} in
this calculation, which would further increase the value of $q$ but
is formally of order $\kappa^2$.

While we take into account realistic weights assigned to each galaxy from
the measurement errors, the results presented here are mainly for illustrative 
purposes: sharp size and flux cuts are rarely made in practice.
Nevertheless, in many cases some form of apparent size cut is applied to 
galaxy catalogs.  
For instance, the main observable of weak lensing studies is galaxy ellipticity, which can only be reliably measured if the galaxies are significantly more extended than the PSF of the
instrument.  Consequently, weak lensing studies usually introduce a size and 
magnitude dependent error weighting function $\eps(m,r)$ which 
operates as a selection function. In general, this selection will
be a smooth function of $m$, $r$, rather than a sharp cut. In that case, 
the resulting
$q$ will be a suitable average over the sharp-cut $q$ values presented here.

Figure \ref{fig:beta} shows the parameters $\beta_f$ and $\beta_r$ as a
function of both the magnitude and size cuts, while holding the other cut
fixed. The corresponding value of $q$ is shown (solid line) as well as the $q$
obtained when neglecting size bias (long-dashed line).  It is clear from the
figure that size bias can have a significant impact on $q$ for size cuts larger
than the instrumental PSF, eventually dominating the lensing effect for large
cuts.   The fact that size cuts smaller than the PSF have no impact on the
total lensing bias is not surprising: since all galaxies have an apparent size
at least as large as the PSF, there is effectively no size cut if $\rcut$ is
smaller than the PSF.  We conclude that the sample of source galaxies in the
DES weak lensing survey will be affected by size bias at a level that is
comparable to or larger than the corresponding magnification bias. 
Interestingly, we find that for typical cut values, the $q$ parameter
including size bias is positive, while it is negative when no size cut
is applied. It might be possible to confirm this prediction with
existing data from galaxy surveys.
As shown in the companion paper \cite{paperII} and briefly in
Sec.~\ref{sec:shear}, lensing bias will turn out to be significant when
interpreting the results from upcoming cosmic shear experiments such as DES.

We have also investigated whether size bias could enhance the lensing signature in the Luminous Red Galaxy (LRG) sample in the Sloan Digital Sky Survey, and in particular in the correlation function of these galaxies. The sample does 
include an effective size cut in order to separate stars from galaxies.  However, since these galaxies are quite bright and large, most of them lie far from the thresholds.  Therefore, size bias
plays an insignificant role in the interpretation of
the LRG correlation function. We estimated that this will also hold for 
higher-redshift LRG samples derived from the SDSS.
For a different reason, size bias is expected to be unimportant for 
quasars (QSO): since these are point-like objects, lensing does not affect
their size noticeably, so that QSO are again only affected by magnification
bias.

\vspace*{-0.27cm}
\section{Effect of lensing bias on shear measurements}
\label{sec:shear}

As an application of our discussion of lensing bias, we briefly present
the effect on shear observables which is the subject of \cite{paperII}. 
Since the cosmic shear field $\gamma$
can be measured only where there are background galaxies, $\gamma$ is
preferentially sampled in regions with a high density of background galaxies.
However, a part of the fluctuations in the galaxy density come from
lensing bias which is due to the \textit{same lensing field} as the shear
$\gamma$ itself. Because of this, the actual measured shear field is 
schematically given by (see \cite{paperII} for details):
\be
\gamma_{\rm obs}(\vth) =[1 + q\,\kappa(\vth)]\,\gamma(\vth).
\ee
The leading lensing bias correction to the shear power spectrum is formally 
identical to the reduced shear correction \cite{DodelsonEtal,Shapiro}. 
However, the amplitude of the correction is multiplied by a 
factor $1+q \approx 2-3$, from our estimates in Sec.~\ref{sec:forecast}.
Thus, the correction to the shear power spectrum from lensing bias is at 
the 5\% level for  $\l\sim1000$, and growing for higher $\l$ \cite{paperII}. \\

\vspace*{-0.8cm}
\section{Summary and Conclusions}

\vspace*{-0.2cm}
Gravitational lensing magnifies galaxies, making them not only brighter, but also larger in appearance.   If galaxy selection criteria include any form of size measure, it follows that gravitational lensing can scatter galaxies in and out of the sample across the applied size cut. This effect, which we have dubbed {\it size bias}, is completely analogous to magnification bias, and is important whenever a size cut is applied that removes some non-negligible fraction of galaxies from the sample. Size bias will often be just as important as magnification bias, so many previous examples subject to the latter might profitably be re-examined for traces of size bias. Our calculations in \S{III} suggest that upcoming large galaxy surveys will be similarly afflicted by both types of lensing bias. 
In a companion paper \cite{paperII}, we evaluate the impact of lensing bias on 
shear observables, and demonstrate that this effect is significant at the 
$5\%$ level for $l\sim 1000$ in case of the shear power spectrum. We estimate
that for DES, neglecting lensing and size bias can lead to biases in the
Dark Energy parameters estimated from cosmic shear at the $2-3\sigma$ level.
Similar conclusions are expected to apply to other upcoming surveys such as
Pan-STARRS \cite{PanStarrs}.

\vspace*{-0.5cm}
\acknowledgments 
\vspace*{\sectspace} 
This work was supported in part by the Kavli Institute for Cosmological Physics at the University of Chicago through grants NSF PHY-0114422 and NSF PHY-0551142. ER was funded by the Center for Cosmology and Astro-Particle Physics (CCAPP) at The Ohio State
University, and by NSF grant AST 0707985. SD is supported by the US Department of Energy including grant 
DE-FG02-95ER40896. LH is supported by DOE grant DE-FG02-92-ER40699, and
ES is supported by grant DE-AC02-98CH10886.

\vspace*{-0.5cm}
\bibliography{sizebias}

\end{document}